\documentstyle[twocolumn,aps,epsf]{revtex}
\tighten
\begin{document}
\draft
\title{Triplet superconductivity in a one-dimensional ferromagnetic 
$t-J$ model}

\author{G.I. Japaridze$^1$ and E. M\"uller-Hartmann}

\address{
Institut f\"ur theoretische Physik,\\ 
Universit\"at K\"oln, 50937 K\"oln, Germany}

\address{~
\parbox{14cm}{\rm 
\medskip
In this paper we study the ground state phase diagram of a one-dimensional 
$t-U-J$ model, at half-filling. In the large-bandwidth limit and for
ferromagnetic exchange with easy-plane anisotropy, a phase with gapless 
charge and massive spin excitations, characterized by the coexistence of
triplet superconducting ($TS$) and spin density wave ($SDW^{z}$) instabilities 
is realized in the ground state. With reduction of the bandwidth, a transition 
into an insulating phase showing properties of the spin-$\frac{1}{2}$ $XY$ 
model takes place. In the case of weakly anisotropic antiferromagnetic 
exchange the system shows a long range dimerized (Peierls) ordering in the 
ground state. The complete weak-coupling phase diagram of the model, including 
effects of the on-site Hubbard interaction, is obtained.  
 }}
\maketitle
\narrowtext

\maketitle

\section{\bf Introduction}

Soon after the discovery of superconductivity in copper-oxide systems, 
a new oxide-superconductor, $Sr_{2}RuO_{4}$, was discovered 
\cite{Trexp1}. Having the same layered perovskite structure as 
$La_{2}CuO_{4}$ the layered ruthenate shows a rather unconventional 
superconducting phase \cite{Trexp2,Trexp3,Trexp4}. Shortly after the discovery 
of $Sr_{2}RuO_{4}$ it was suggested that a triplet superconducting 
phase is realized in this compound \cite{RS,Baskaran,Machida}. Since then 
convincing experimental evidence has been collected that $Sr_{2}RuO_{4}$ is 
most likely a $p$-wave superconductor (for a recent review see 
\cite{Sigrist}). An important feature of related ruthenate compounds is close 
proximity to magnetic instability ($SrRuO_{3}$ and  $Sr_{2}RuYO_{6}$ are 
ferro- and antiferromagnetic, respectively), indicating strong correlations 
in the $Ru$ ions. The NMR studies clearly show tendency towards 
ferromagnetism in $Sr_{2}RuO_{4}$ \cite{Trexp4pr}. Moreover, 
very recent experiments indicate the easy-plane anisotropy of ferromagnetic 
spin fluctuations in this compound \cite{Trexp5}. The presence of 
ferromagnetism in $SrRuO_{3}$ and the analogy with $^{3}He$ made Sigrist and 
Rice predict the triplet nature of superconductivity in $Sr_{2}RuO_{4}$ 
\cite{RS}. Close proximity of the ferromagnetic and triplet superconducting 
instabilities in $Sr_{2}RuO_{4}$ increase the interest in models providing a
mechanism for Cooper pairing via ferromagnetic spin fluctuations 
\cite{RS,MS,Sigrist}.

Another group of unconventional superconductors showing close proximity of 
magnetic  and superconducting ordering belongs to the $(TMTSF)_{2}X$ family 
of quasi-one-dimensional conductors (the Bechgaard salts)\cite{Jerome1}. At 
ambient pressure, most of these compounds show a spin-density wave ($SDW$) 
ordering in the ground state. Under moderate pressure, the $SDW$ instability 
is suppressed and replaced by a superconducting transition at a critical 
temperature of the order of 1 K \cite{Jerome2}. The most interesting 
exceptions to this scheme are: 1) $(TMTSF)_{2}ClO_{4}$ which is 
supperconducting at ambient pressure and 2) $(TMTSF)_{2}PF_{6}$ which shows a 
spin-Peierls ($SP$) phase in the ground state at atmospheric pressure. In this 
latter case, increasing pressure leads first to a transition from the $SP$ 
phase into a $SDW$ phase, and finally to the suppression of the $SDW$ 
ground state in favor of superconductivity
\cite{Jerome2}. Triplet superconducting ordering in Bechgaard salts was 
suggested soon after the discovery of $TMTSF$ systems \cite{AbrikosGorkov} to 
explain the strong suppression of $T_{c}$ by nonmagnetic impurities 
\cite{Trexp6}. Although the symmetry of the supercunducting phase in 
Bechgaard salts still remains the subject of some controversy, growing 
experimental evidence has been collected in the last few years, indicating 
that 
the Bechgaard salts $(TMTSF)_{2}ClO_{4}$ and $(TMTSF)_{2}PF_{6}$ under 
pressure are triplet superconductors ($TS$) \cite{Trexp8}.  

In this paper we put forward a rather simple extension of the Hubbard model by
incorporating direct anisotropic exchange (of either sign) between electrons 
on nearest-neighbor sites. In 1D the Hamiltonian reads:
\begin{eqnarray}\label{tJmodel} 
{\cal H}& = & -t\sum_{n,\alpha}(c^{\dagger}_{n,\alpha}
c^{\vphantom{\dagger}}_{n+1,\alpha}
+c^{\dagger}_{n+1,\alpha}c^{\vphantom{\dagger}}_{n,\alpha})\nonumber\\
& +&  
U \sum_{n}c^{\dagger}_{n,\uparrow}c^{\vphantom{\dagger}}_{n,\uparrow}
c^{\dagger}_{n,\downarrow}c^{\vphantom{\dagger}}_{n,\downarrow}\nonumber\\
&+&\sum_{n}\{ \frac{1}{2}J_{\perp}(S^{+}_{n}S^{-}_{n+1} + h.c.)+
J_{\parallel}S^{z}_{n}S^{z}_{n+1}\}
\end{eqnarray}
Here $c^{\dagger}_{n,\alpha}$ ($c^{\vphantom{\dagger}}_{n,\alpha}$) is the 
creation (annihilation) operator for an electron at site $n$ with spin 
${\alpha}$, $\vec {\bf S}(n)= \frac{1}{2}c^{\dagger}_{n,\alpha}\vec
\sigma^{\vphantom{\dagger}}_{\alpha\beta}c^{\vphantom{\dagger}}_{n,\beta}$ 
where $\sigma^{{\it i}}$ (${\it i}=x,y,z$) are the Pauli matrices.

The model (\ref{tJmodel}) was intensively studied in the context of 
High-$T_{c}$ superconductivity for strong on-site repulsion and for
isotropic antiferromagnetic exchange \cite{DRS,RTT,Fabrizio,PSW}. 
Below we study the weak-coupling phase diagram of the model (\ref{tJmodel}) 
focusing on effects of exchange anisotropy, in particular in the case 
of ferromagnetic exchange.  We will show that the one-dimensional version of 
this (\ref{tJmodel}) model has a ground state phase diagram characterized by 
the close proximity of {\em triplet superconducting}, {\em spin density wave}, 
{\em ferromagnetic} and {\em Peierls dimerized} phases.

That the $TS$ phase can be realized in 1D correlated electron systems is well
known from standard ``{\em g-ology}'' studies \cite{Solyom}. 
The extended ($U$-$V$) Hubbard model with nearest-neighbor attraction ($V<0$) 
has been intensively studied to explain the competition between $SDW$ and 
superconducting instabilities in $TMTSF$ compounds \cite{Fukuyama}. However,
due to spin rotational invariance, in the extended Hubbard model the $TS$ 
phase is realized only in the Luttinger liquid phase for
$|U|<-2V$ \cite{Solyom,Emery,Voit}, where both charge and
spin excitations are gapless. Singlet superconducting ($SS$) and $TS$ 
correlations show identical power-low decay at large distances and the $TS$ 
instability dominates only due to weak logarithmic corrections \cite{Voit}. 
On the other hand, in the spin gapped phase $U<2V$, the dynamical 
generation of a spin gap leads to the {\em complete suppression} of the $TS$ 
and $SDW$ instabilities. 

In this paper we study the weak-coupling ground state phase diagram of the 
model (\ref{tJmodel}) at half-filling. As we will show below, in the case of  
{\em ferromagnetic easy-plane anisotropy} 
($J_{\perp} < J_{\parallel} < 0$), the $TS$ and $SDW$ are the only 
instabilities in the system. In some sense, the {\em ferromagnetic} 
$t-J$ model (\ref{tJmodel}) shows infrared behavior which is dual to that of 
the attractive $U-V$ Hubbard model. This duality is most easily seen by
comparing the attractive Hubbard model ($U < 0, J_{\parallel}= J_{\perp} = 0$) 
and the {\em ferromagnetic} itinerant $XY$ model 
($J_{\perp} < 0, U = J_{\parallel} = 0$). In both models the spin excitation 
spectrum is gapped and the charge excitation spectrum is gapless. In the 
attractive Hubbard model the dynamical generation of the spin gap is associated 
with the suppresion of $SDW$ and $TS$ fluctuations. In the ground state only 
the charge density wave ($CDW$) and $SS$ correlations survive. At half-filling, 
due to $SU(2)$-symmetry of the charge channel, the ground state is 
characterized by the coexistence of $CDW$ and $SS$ instabilities. Away from 
half-filling the singlet superconducting instability dominates \cite{FK}. In 
the (weak-coupling limit of the) ferromagnetic itinerant $XY$ model, however, 
due to the $U(1)$-spin symmetry, the dynamical generation of a spin gap leads 
to {\em complete suppression} of the $SS$ and $CDW$ fluctuations. At 
half-filling, due to the $SU(2)$-symmetry of the charge channel, the $TS$ and 
$SDW$ instabilities coexist (see Fig. 1). Doping of the system, as in the case 
of the Hubbard model, splits the degeneracy, in this case in favor of the 
$TS$ ordering.
  
The Ising part of the ferromagnetic exchange tends to reduce the $TS$ ordering. 
The line $J_{\parallel} = J_{\perp} = J < 0$, corresponding to {\em isotropic 
ferromagnetic exchange} marks the transition into a regime with gapless spin 
excitations. At half-filling, in the case of isotropic exchange the model 
(\ref{tJmodel}) is characterized by the high $SU(2) \otimes SU(2)$ symmetry. 
Due to this symmetry the ground state of the ferromagnetic itinerant $XXX$ 
model is a Luttinger liquid ($LL$) phase, characterized by an identical 
power-law decay of all correlations at large distances. In the case of 
{\em ferromagnetic easy-axis} anisotropy ($J_{\parallel} < J_{\perp} < 0$) 
a $LL$ phase with weakly dominating easy-plane magnetic instabilities is 
realized. In the case of antiferromagnetic exchange, the line 
$J_{\perp}= -\frac{1}{2}J_{\parallel}$ marks the transition into a regime 
where a charge gap opens. Therefore in the case of {\em antiferromagnetic 
easy-axis} anisotropy ($J_{\parallel} > 2|J_{\perp}|$) both the charge and 
the spin channels are massive and long range $SDW^{z}$ (N\'eel) ordering 
takes place. 

\begin{figure}
\mbox{\epsfxsize 8.0cm\epsffile{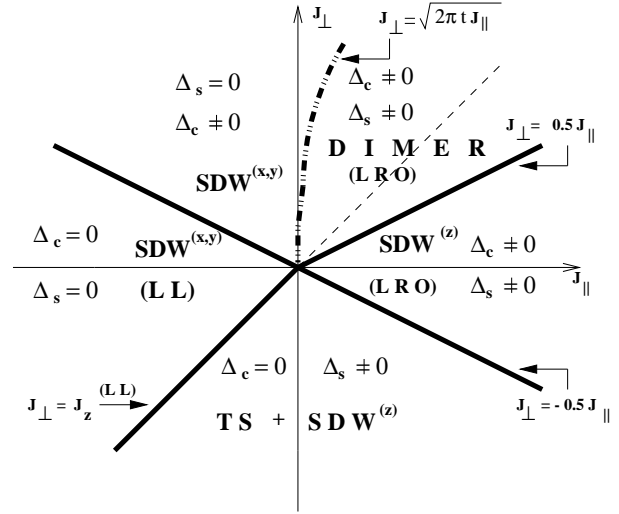}}
\vskip0.5cm
\caption[]{
The weak-coupling phase diagram of the model (\ref{tJmodel}) in the case of a
half-filled band and at $U=0$. $\Delta_{c(s)}$ denotes the charge (spin) gap. 
Thick lines seperate different phases:  1. $Dimer$ (LRO) - long range 
ordered dimerized (Peierls) phase. 2. $SDW^{z}$ (LRO) - the long range ordered 
antiferromagnetic (N\'eel) phase. 3. $SDW^{x,y}$ - insulating state with 
dominating easy-plane antiferromagnetic correlations. 4. $SDW^{x,y}$  (LL)
- Luttinger liquid phase with dominating easy-plane antiferromagnetic 
correlations. 5. $TS+SDW^{z}$ - phase with gapless charge and gapped spin 
excitation spectrum characterized by the coexistence of the triplet 
superconducting and antiferromagnetic instabilities.}
\label{fig1}
\end{figure} 

Very rich is the phase diagram of the model (\ref{tJmodel}) in the case of 
{\em antiferromagnetic} exchange. The line 
$J_{\perp} =\frac{1}{2}J_{\parallel} > 0$ is the transition line from the 
LRO $SDW^{z}$ phase into a LRO dimerized 
(Peierls) phase.  The long range ordered dimerized phase is realized in 
particular in the ground state of the antiferromagnetic itinerant $XXX$ model 
($J_{\parallel} = J_{\perp} > 0$). The line $J_{\perp} =\sqrt{2\pi t 
J_{\parallel}}$ marks the transition into an insulating phase with 
gapless spin excitation spectrum and dominating in-plane ($XY$) magnetic 
correlations.

We have to stress the weak-coupling nature of the presented phase diagram. 
Higher order corrections will modify the shape of borderlines between phases. 
However, far more important are strong coupling effect. In the case of 
strong exchange interaction, there are additional phase transitions due to the
finite band width. Usually such effects can not be traced within the 
continuum-limit (infinite band) approach used in this paper and will require 
numerical studies. Below we focus only on the $TS$ part of the phase diagram 
and present a qualitative analysis of the transition from the $TS$ phase 
into a magnetic insulating phase.

Let us first consider the itinerant $XY$ model ($J_{\parallel}=U=0$). In the 
weak-coupling limit $|J_{\perp}| \ll t$, the charge excitation spectrum is 
gapless and the spin excitation spectrum is massive. However, in the limit of 
strong ferromagnetic exchange $|J_{\perp}| \gg t$, the model is equivalent to 
the $XY$ spin chain. Therefore, with increasing coupling one has to expect a 
transition from the regime with massive spin and massless charge excitation
spectrum into a insulating magnetic phase with gapless spin excitations.
Our finite system studies show (see Fig. 2) that this transition takes place at 
$J^{c}_{\perp} \sim -4t$ and is of level crossing type \cite{Note}. After 
the transition the ground state energy of the itinerant model becomes very 
close to the ground state of the spin-$\frac{1}{2}$ $XY$ chain. In the case of 
antiferromagnetic exchange 
there is no transition with increasing $J_{\perp}>0$ and the system 
continuously approaches its limiting behavior at 
$J_{\perp}/t \rightarrow \infty$. 
\begin{figure}
\mbox{\epsfxsize 8.0cm\epsffile{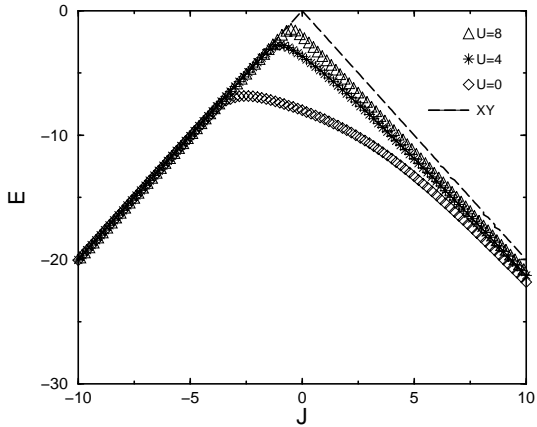}}
\vskip0.5cm
\caption[]{The ground state energy of the half-filled itinerant $XY$-Hubbard 
chain (6 sites) vs. exchange for $U=0$ (diamonds ), $U=4$ (stars ) and $U=8$ 
(triangles). The dashed line correspond to the ground state energy of the 
spin-$\frac{1}{2}$ $XY$ model.}
\label{fig2}
\end{figure} 

The numerical data presented in Fig. 2 clearly indicate the
renormalization of the critical value of the transverse exchange 
$J^{c}_{\perp}$ by the on-site Hubbard interaction. In the limit of strong 
Hubbard repulsion $J^{c}_{\perp}$ is reduced to values of the order  
$t^{2}/U$. Detailed numerical studies of the strong-coupling phase diagram of 
the model (\ref{tJmodel}) is in progress and will be published elsewhere.

Figure 3 shows phase diagram of the itinerant-$XY$-Hubbard model. 

Below we will focus on the ferromagnetic part of the phase diagram. We will 
see that for moderate values of the Hubbard repulsion the 
$TS$ and the $SDW$ phases survive. In the case of weak exchange one obtains  
that a charge gap opens at $U >-J_{\perp}$ and a transition into a 
long range ordered $SDW^{z}$ phase takes place. Therefore, at $U > 0$ with 
increasing exchange one has to expect two different transitions: 
for $U<|J_{\perp}|<t$ the transition discussed above will take place, but 
for $U \gg t, |J_{\perp}|$ a ``spin-flop'' transition from the 
LRO $SDW^{z}$ phase into the $XY$ phase has to occur. 

\begin{figure}
\mbox{\epsfxsize 8.0cm\epsffile{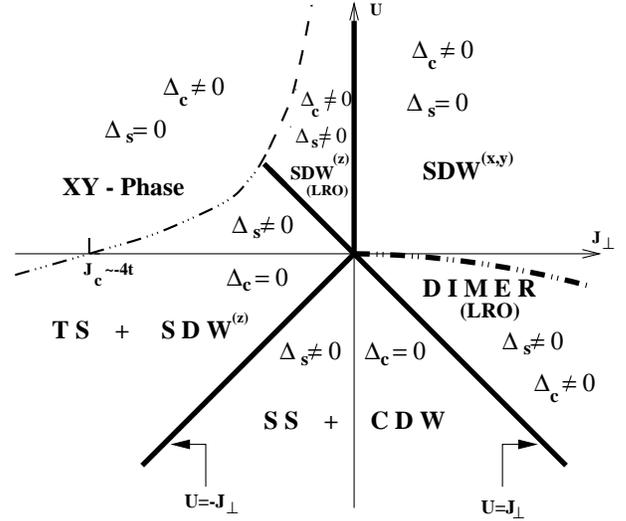}}
\vskip0.5cm
\caption[]{
The weak-coupling phase diagram of the model (\ref{tJmodel}) 
at $J_{\parallel}=0$. Solid lines indicate borders between the weak-coupling 
limit phases. The dashed line marks (qualitatively) the transition 
into the $XY$ magnetic phase.}
\label{fig3}
\end{figure} 

The paper is orginized as follows: in the next section the weak-coupling 
continuum-limit version of the model (\ref{tJmodel}) is constructed and the 
renormalization-group analysis is performed. In the Sect. III, the 
weak-coupling phase diagram is discussed. Finally, Sect. IV is devoted to a 
discussion and to concluding remarks.

\section{\bf Continuum-limit theory and bosonization.}

In this section we construct the continuum-limit version of the model 
Eq.~(\ref{tJmodel}) at half-filling. While this procedure has a long history 
and is reviewed in many places \cite{GNT}, for clarity we briefly sketch the 
most important points.

The field theory treatment of 1D systems of correlated electrons is based on 
the weak-coupling approach $|U|, |J_{\perp}|, |J_{\parallel}| \ll t$. Assuming 
that the low energy physics is controlled by states near the Fermi points 
$ \pm k_{F}$ ($k_{F} =\pi/2a_{0}$, where $a_{0}$ is the lattice spacing) we 
linearize the spectrum around these points and obtain two species (for each 
spin projection $\alpha$) of fermions, $R_{\alpha}(n)$ and $L_{\alpha}(n)$, 
which describe excitations with dispersion relations $E = \pm v_{F}p$. Here, 
$v_{F}=2 ta_{0}$ is the Fermi velocity and the momentum $p$ is measured from 
the two Fermi points. More explicitly, one decomposes the momentum expansion 
for the initial lattice operators into two parts centered around $\pm k_{F}$ 
to obtain the mapping:
\begin{equation}\label{linearization}
c_{n,\alpha} \rightarrow  {\it i}^{n}R_{\alpha}(n) + 
(-{\it i})^{n}L_{\alpha}(n),
\end{equation}
where the fields $R_{\alpha}(n)$ and $L_{\alpha}(n)$ describe 
right-moving and left-moving particles, respectively, and are assumed to be 
smooth on the scale of the lattice spacing. This allows us to introduce the 
continuum fields $R_\alpha(x)$ and $L_\alpha(x)$ by
\begin{eqnarray}\label{continuumfields}
R_{\alpha}(n) & \rightarrow & \sqrt{a_{0}}R_{\alpha}(x=na_{0}),\nonumber\\
L_{\alpha}(n) & \rightarrow & \sqrt{a_{0}}L_{\alpha}(x=na_{0}).
\end{eqnarray}

In terms of the continuum fields the free Hamiltonian reads:
\begin{equation}\label{freelinearized}
{\cal H}_{0}  =  E_{0} - iv_{F}\sum_\alpha \int  dx 
[:R^{\dagger}_{\alpha}\partial_{x} R_{\alpha}: - 
:L^{\dagger}_{\alpha}\partial_{x}L_{\alpha}:]
\end{equation}
which is recognized as the Hamiltonian of a free massless Dirac field and the 
symbols :...: denote normal ordering with respect to the ground state of the 
free system.

The advantage of the linearization of the spectrum is twofold: the initial 
lattice problem is reformulated in terms of smooth continuum fields and -- 
using the bosonization procedure -- is mapped to the theory of two 
independent (in the weak-coupling limit) quantum sine-Gordon (SG) models 
describing charge and spin degrees of freedom, respectively. 

In terms of the continuum fields the initial lattice operators have the form:
\begin{eqnarray}\label{mapping2}
&\hat\rho_{n,\alpha}-\frac{1}{2}=a_{0}\{(J_{R,\alpha} + J_{L,\alpha})&
\nonumber\\
&+(-1)^{n}(R^{\dagger}_{\alpha}(x)L_{\alpha}(x) +
L^{\dagger}_{\alpha}(x)R_{\alpha}(x))\},&
\end{eqnarray}
here $J_{R,\alpha} \equiv :R^{\dagger}_{\alpha}(x)R_{\alpha}(x):$ and  
$J_{L,\alpha} \equiv :L^{\dagger}_{\alpha}(x)L_{\alpha}(x):$,
\begin{equation}
\vec {\bf S}(n) = a_{0}\cdot\{ \vec {{\bf M}}(x) + 
(-1)^{n}\vec{{\bf L}}(x)\}.
\end{equation}
where
\begin{equation}
\vec {\bf M}(x) = R^{\dagger}_{\alpha}(x) 
\frac{\vec {\sigma}_{\alpha\beta}}{2}R_{\beta}(x) + 
L^{\dagger}_{\alpha}(x)\frac{\vec {\sigma}_{\alpha\beta}}{2}L_{\beta}(x)
\end{equation}
determines the smooth part of the spin density in the continuum limit, and
\begin{equation}
\vec {\bf L}(x) = R^{\dagger}_{\alpha}(x)\frac{\vec {\sigma}_{\alpha\beta}}{2}
L_{\beta}(x) + L^{\dagger}_{\alpha}(x)\frac{\vec {\sigma}_{\alpha\beta}}{2} 
R_{\beta}(x)
\end{equation}
is the staggered part of the local spin density.

The second step is to use the standard bosonization expressions for fermionic 
bilinears \cite{GNT}:
\begin{eqnarray}\label{bos1}
-i\sum_{\alpha} [:R^{\dagger}_{\alpha}\partial_{x} R_{\alpha}:&-& 
:L^{\dagger}_{\alpha}\partial_{x}L_{\alpha}:] \rightarrow \nonumber\\
{1\over 2}\{(\partial_{x}\theta_{c})^2+(\partial_{x}\phi_{c})^2\}
& + &{1\over 2}\{(\partial_{x}\theta_{s})^2+(\partial_{x}\phi_{s})^2\},
\end{eqnarray}
\begin{eqnarray}
J_{R,\alpha}+J_{L,\alpha}&\rightarrow&
{1 \over {\sqrt{2\pi}}}[(\partial_{x}\phi_{c}) + \alpha
(\partial_{x}\phi_{s})],\label{bos2}\\
J_{R,\alpha}-J_{L,\alpha}& \rightarrow & {1 \over {\sqrt{2\pi}}}
[(\partial_{x}\theta_{s}) + \alpha (\partial_{x}\theta_{s})]\label{bos3}\\
R^{\dagger}_{\alpha}(x)R_{-\alpha}(x)&\rightarrow &
{1 \over 2 \pi a_{0}}\exp(- {\it i}\alpha\sqrt{2\pi}(\phi_{s}-\theta_{s}))
\label{bos4}\\
L^{\dagger}_{\alpha}(x)L_{-\alpha}(x)&\rightarrow &
{1 \over 2 \pi a_{0}} \exp(+ {\it i} \alpha \sqrt{2\pi}(\phi_{s} + 
\theta_{s}))\label{bos5}\\
R^{\dagger}_{\alpha}(x)L_{\alpha}(x)&\rightarrow &  
{-i \over 2 \pi a_{0}}\exp(+{\it i} \sqrt{2\pi} 
(\phi_{c} + \alpha \phi_{s})),\label{bos6}\\
R^{\dagger}_{\alpha}(x)L_{-\alpha}(x)&\rightarrow &  
{1 \over 2 \pi a_{0}}\exp(- {\it i} \sqrt{2\pi} 
(\phi_{c} - \alpha \theta_{s})).\label{bos7}
\end{eqnarray}
Here scalar fields $\phi_{c,s}(x)$ describe the charge and the spin degrees of 
freedom and fields $\theta_{c,s}(x)$ are their dual counterparts: 
$\partial_{x}\theta_{c,s}= \Pi_{c,s}$ where $\Pi_{c,s}$ is the momentum 
conjugated to the field $\phi_{c,s}$. 

Using bosonization formulas (\ref{bos1})-(\ref{bos7}), after rescaling of the
fields and lengths, the continuum-limit version of the Hamiltonian 
(\ref{tJmodel}) acquires the following form 
\begin{equation}\label{bosHamiltonian}
{\cal H}= {\cal H}_{c}+{\cal H}_{s}
\end{equation}
where
\begin{eqnarray}
&{\cal H}_{c}=v_{c}\int dx\Big\{{1\over 2}[(\partial_{x}\varphi_{c})^2 
+(\partial_x \vartheta_{c})^2]&\nonumber\\
&+ \frac{m_{c}}{a_0^2}\cos(\sqrt{8\pi K_{c}}\varphi_{c})
\Big\},&\label{SGc}\\             
&{\cal H}_{s}=v_{s}\int dx\Big\{{1\over 2}[(\partial_{x}\vartheta_{s})^2
+(\partial_x\varphi_{s})^2]\nonumber\\
&+ \frac{m_{s}}{a_0^2}\cos(\sqrt{8\pi K_{s}}\varphi_{s})\Big\}.&\label{SGs} 
\end{eqnarray}

Here we have defined
\begin{eqnarray}
& K_{c} \simeq 1+{1\over 2}g_{c},\qquad
m_{c}=-\frac{1}{2\pi}g_{u},&\label{Kc}\\
& K_{s}\simeq 1+{1\over 2}g_{s},\qquad
m_{s}=\frac{1}{2 \pi }g_{\perp},&\label{Ks}
\end{eqnarray}
$v_{c(s)}=v_{F}K^{-1}_{c(s)}$ and  small dimensionless coupling constants
given by:
\begin{eqnarray}
g_{c} =  g_{u}&= &- {1 \over 2 \pi t}(U+J_{\perp}+\frac{1}{2}J_{\parallel}),
\label{gcgu}\\
g_{s} &= & {1 \over 2\pi t}(U + J_{\perp}-\frac{3}{2}J_{\parallel} ),
\label{gperp}\\
g_{\perp}  & =  &{1 \over 2\pi t}(U - J_{\perp}+\frac{1}{2}J_{\parallel} ).
\label{gs}
\end{eqnarray}

The relation between $K_{c}$ ($K_{s}$), $m_{c}$ ($m_{s}$), and $g_{c}$ 
($g_{s}$), $g_{u}$ ($g_{\perp}$) is universal in the 
weak coupling limit. In obtaining (\ref{bosHamiltonian}), several terms 
corresponding to scattering processes in the vicinity of a Fermi point, which 
lead to a renormalization of the Fermi velocities in second order in $g$, as 
well as strongly irrelevant terms $\sim \cos(\sqrt{8\pi K_{c}}\varphi_{c})
\cos(\sqrt{8\pi K_{s}}\varphi_{s})$ describing um\-klapp processes with 
parallel spins, were omitted. 
 
The mapping of the initial lattice Hamiltonian Eq.~(\ref{tJmodel}) into 
the continuum theory of two decoupled quantum SG models
Eqs.~(\ref{SGc})-(\ref{SGs}) performed above allows the study the ground state 
phase diagram of the system based on the infrared properties of the SG 
Hamiltonians. The corresponding behavior of the SG model is described by 
pairs of renormalization group equations for the effective coupling constants 
$\Gamma_{i}$ \cite{Wieg}  
\begin{eqnarray}
d\Gamma_{u}/dL &=& - \Gamma_{c} \Gamma_{u} , \qquad 
d\Gamma_{c}/dL = - \Gamma^{2}_{u},\label{RGsg.a}\\ 
d\Gamma_{\perp}/dL & = &- \Gamma_{s} \Gamma_{\perp},\qquad 
d\Gamma_{s}/dL  = - \Gamma^{2}_{\perp},\label{RGsg.b}  
\end{eqnarray}
where $L=log{\left(a/a_{0}\right)}$ and $\Gamma_{i}(0) = g_{i} $. Each pair 
of equations 
(\ref{RGsg.a}) and (\ref{RGsg.b}) describes a Kosterlitz--Thouless 
transition \cite{KT} in the charge and spin channels. The flow diagram is 
given in Fig. 1. The flow lines lie on the hyperbola  
\begin{equation}\label{flowlines} 
\Gamma_{c (s)}^{2} - \Gamma_{u (\perp)}^{2} = \mu_{c (s)} = 
g_{c (s)}^{2} - g_{u (\perp)}^{2}, 
\end{equation} 
and -- depending on the relation between the bare coupling constants 
$g_{c(s)}$ and $g_{u(\perp)}$ -- exhibit two different regimes:


\begin{figure}
\mbox{\epsfxsize 8.0cm\epsffile{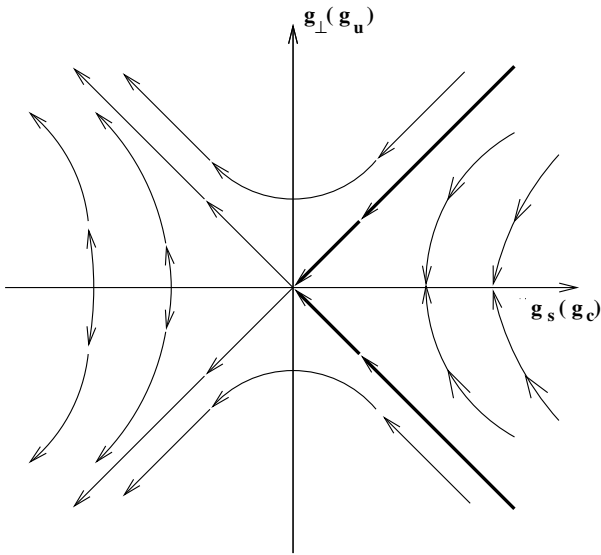}}
\vskip0.5cm
\caption[]{The renormalization-group flow diagram; the arrows denote the 
direction of flow with increasing length scale.}
\label{TRSCsgsd.eps}
\end{figure} 

For $g_{c}\geq |g_{u}|$ ($g_{s}\geq |g_{\perp}|$) we are in the weak coupling 
regime; the effective mass $M_{c (s)}$ scales to $0$. The 
low energy (large distance) behavior of the gapless charge (spin) degrees of 
freedom is described by a free scalar field
\begin{equation}\label{KG}
{\cal H}_{c (s)} ={1 \over 2} v_{c (s)} \int dx 
\{(\partial_{x} \theta_{c (s)})^{2} + 
(\partial_{x} \varphi_{c (s)})^{2}\} 
\end{equation}
where $\partial_{x}\theta_{c (s)}=P_{c (s)}$.

The corresponding correlations show a power law decay
\begin{eqnarray}
\langle e^{i\sqrt{2\pi K}\varphi(x)} e^{-i \sqrt{2\pi K} \varphi (x')} 
\rangle \sim \left| x - x' \right|^{- K},\label{freecorrelations1}\\
\langle e^{i \sqrt{2\pi/K}\theta(x)} e^{-i\sqrt{2\pi/K}
\theta(x')}\rangle 
 \sim \left| x - x' \right| ^{-1 / K},\label{freecorrelations2}
\end{eqnarray}
and the only parameter controlling the infrared behavior in the gapless 
regime is the fixed-point value of the effective coupling constants 
$K_{c(s)}$. 

For $g_{c}<|g_{u}|$ ($g_{s}<|g_{\perp}|$) the system scales to the strong 
coupling regime; depending on the sign of the bare mass $m_{c(s)}$ the 
effective mass $M_{c(s)}$ scales to $\pm\infty$, which signals the crossover to 
the strong coupling regime and indicates the dynamical generation of a 
commensurability gap in the charge (spin) excitation spectrum. The fields 
$\varphi_{c}$ ($\varphi_{s}$) get ordered with the vacuum expectation values 
\cite{ME}  
\begin{equation}\label{orderfields}
\langle\varphi_{c(s)}\rangle =\left\{ \begin{array}{l@{\quad}} 
\sqrt{\pi/8K_{s}}
\hskip0.5cm(m_{c(s)}>0) \\ 
0 \hskip1.7cm (m_{c(s)}<0)\end{array}\right. \, . 
\end{equation} 

Using the initial values of the coupling constants, given in
(\ref{gcgu})-(\ref{gs}), we see that flow trajectories in the charge sector 
(due to the $SU(2)$-charge symmetry) are along the separatrix $g_{c}= g_{u}$. 
Therefore, at 
\begin{equation}\label{CSsc}
U+J_{\perp}+\frac{1}{2}J_{\parallel}>0.
\end{equation}
there is a gap in the charge excitation spectrum ($\Delta_{c} \neq 0$) and the 
charge field $\varphi_{c}$ is ordered  with the vacuum expectation value 
\begin{equation}\label{PHIcs}
\langle  \varphi_{c} \rangle  = 0, 
\end{equation}
while at $U+J_{\perp}+\frac{1}{2}J_{\parallel}<0$ the charge sector is gapless 
and the fixed-point value of the parameter $K^{\ast}_{c}$ is $1$.

The $U(1)$ symmetry of the spin channel ensures more alternatives. Depending 
on the relation between the bare coupling constants there are {\bf 
two different strong-coupling sectors in the spin channel}. For 
\begin{equation}\label{SS1sc}
U < \frac{1}{2}J_{\parallel} < J_{\perp}-U
\end{equation}
the spin channel is massive ($\Delta_{s} \neq 0$) and the field $\varphi_{s}$ 
gets ordered with the vacuum expectation value 
\begin{equation}\label{PHI1ss}
\langle  \varphi_{s} \rangle  = 0, 
\end{equation}
while for 
\begin{equation}\label{Ss1cc}
J_{\perp}< {\em min}\{U+\frac{1}{2}J_{\parallel}; J_{\parallel}\}
\end{equation}
the spin channel is massive ($\Delta_{s} \neq 0$), with the vacuum 
expectation value
\begin{equation}\label{PHI2ss}
\langle  \varphi_{s} \rangle  = \sqrt{\pi/8K_{s}}.
\end{equation}

In all other cases the excitation spectrum in the corresponding channel is 
gapless. The low-energy behavior of the system is controlled by the fixed-point 
value of the Luttinger-liquid parameter 
$K^{\ast}_{s}=1+\frac{1}{2}g^{\ast}_{s}$. 

However, in the particular case of strong antiferromagnetic easy-plane 
anisotropy ($J_{\perp} \gg |J_{\parallel}|$), the clarification of details of 
the phase diagram requires a closer inspection. Let us first consider the 
$XY$ limit of the model: $U=J_{\parallel}=0$. As we see, the initial values 
of the coupling constants $g_{s}$ and $g_{\perp}$, given in 
(\ref{gperp})-(\ref{gs}), lie exactly on the separatrix $g_{s}=-g_{\perp}$ and 
scale to the $SU(2)$-symmetric fixed-point value $g_{s}=g_{\perp}=0$. However, 
due to the low $U(1)$-symmetry of the model, there is no symmetry reason which 
would guarantee that the bare couplings lie {\em exactly} on the separatrix. 
As we will show below, the higher order (finite band) effects push the scaling 
trajectories from the separatrix. For details of the method we refer the reader 
to the paper \cite{AM} where a similar effect in the pair-hopping model was 
considered.

Since in first order the couplings lie on the separatrix, we must work 
to ${\Large O}(J^{2})$. We find, that in the $SU(2)$-symmetric case 
($J_{\perp}=J_{\parallel}=J$) $g_{s}-|g_{\perp}|=0$ up to 
${\Large O}\left(J^{2}\right)$, but 
for $J_{\perp} \neq J_{\parallel}$ there is an 
${\Large O}\left((J_{\perp}-J_{\parallel})^{2}\right)$ correction to this 
quantity. This correction occurs due to the nonlocal character of the 
interaction and to deviations from the linear dispersion relations for 
electrons on the lattice. 

In particular, upon integrating out all modes with momenta outside the 
small region around each Fermi point $|p-p_{F}| < \Lambda \equiv 2/a$, where 
$\Lambda $ is small compared to $p_{F}$, we obtain the effective 
theory described by equations (\ref{SGc})-(\ref{SGs}), with the following 
coupling constants:
\begin{eqnarray}\label{Gc}
g_{c} & = & g_{u} = -\frac{1}{2 \pi t}(J_{\perp}+
\frac{1}{2}J_{\parallel})  + \frac{1}{(2\pi t)^{2}}
\left(J_{\perp}J_{\parallel}-{1 \over 4}J_{\parallel}^{2}\right)
\nonumber\\
&-& \frac{1}{(2\pi t)^{2}}\left(J_{\perp} + 
\frac{1}{2}J_{\parallel}\right)^{2}log{\left(a/a_{0}\right)} ,\\
g_{\perp} &=& \frac{1}{2\pi t}\left(\frac{1}{2}J_{\parallel} - J_{\perp}\right)
-\frac{1}{(2\pi t)^{2}}\left(J_{\perp}^{2}-2J_{\perp}J_{\parallel}+
\frac{1}{4}J_{\parallel}^{2}\right)\nonumber\\ 
&-& \frac{1}{(2\pi t)^{2}}\left(J_{\perp} - \frac{3}{2}J_{\parallel}\right)
\left(\frac{1}{2}J_{\parallel}- J_{\perp}\right)log{\left(a/a_{0}\right)}
\label{Gperp2order}\\
g_{s} &=& \frac{1}{2\pi t}\left(J_{\perp}-\frac{3}{2}J_{\parallel}\right)
+ \frac{1}{(2\pi t)^{2}}\left(2J_{\perp}^{2}-J_{\perp}J_{\parallel}-
\frac{1}{2}J_{\parallel}^{2}\right)\nonumber\\
&-&\frac{1}{(2\pi t)^{2}}\left(J_{\perp}-\frac{1}{2}J_{\parallel}\right)^{2}
log{\left(a/a_{0}\right)}
.\label{Gs2order} 
\end{eqnarray}
 
Thus we see, while the bare couplings $g_{c}$ and $g_{u}$ always lie on the 
$SU(2)$ separatrix, $g_{s}$ equals $|g_{\perp}|$ only in the $SU(2)$-symmetric 
case $J_{\perp}=J_{\parallel}=J$:
\begin{equation}\label{SSSU2}
g_{s} = g_{\perp} = - \frac{J}{4\pi t}  
-\frac{J^{2}}{4(2\pi t)^{2}}\left[log{\left(a/a_{0}\right)}
- 3\right].
\end{equation}
In the case of the $XY$ model the corresponding parameters are:
\begin{eqnarray}\label{XYgperp}
g_{\perp} & = & - \frac{1}{2\pi t}J_{\perp}+ \frac{1}{(2\pi t)^{2}} 
J_{\perp}^{2}\left[log{\left(a/a_{0}\right)}-1\right]\\
g_{s} & = & \frac{1}{2\pi t}J_{\perp} - 
\frac{1}{(2\pi t)^{2}} J_{\perp}^{2}\left[log{\left(a/a_{0}\right)}
-2\right].\label{XYgs}
\end{eqnarray}
Thus we see that the values of the coupling constants $g_{\perp}$ and 
$g_{s}$ move off the separatrix into the region of the flow diagram that flows 
to non-zero $g_{s}$. This movement occurs because now 
\begin{equation}
\mu = g_{s}^{2}-g_{\perp}^{2}=\left(J_{\perp}/2\pi t\right)^{3} + 
{\Large O}\left(J_{\perp}/t\right)^{4}>0. \label{mu}
\end{equation}
To actually find the 
end-point value of the parameter $\Gamma_{s}$, we need the second order 
renormalization group equations for the effective coupling constants.
These equations read \cite{AmitGrinstein} 
\begin{eqnarray}\label{RGsg2ord.a} 
d\Gamma_{\perp}/dL & = &- \Gamma_{s} \Gamma_{\perp}-\frac{1}{2}
\Gamma^{3}_{\perp}, \nonumber \\ 
d\Gamma_{s}/dL& = & - \Gamma^{2}_{\perp}-
\frac{1}{2}\Gamma^{2}_{\perp}\Gamma_{s}.\label{RGsg2ord.b}  
\end{eqnarray}

Combing equations (\ref{RGsg2ord.a})-(\ref{mu}) one obtains
\begin{equation}\label{RGmu1}
d\mu/dL = -\Gamma_{\perp}^{2}\mu.
\end{equation}
Substituting the first-order solution for $\Gamma_{\perp}$ and solving 
(\ref{RGmu1}) we obtain
\begin{equation}\label{RGmu2}
\mu(\infty) \equiv (g^{\ast}_{s})^{2} = \left(\frac{J_{\perp}}{2\pi t}\right)^{3}\exp(-J_{\perp}/2\pi t).
\end{equation}
At small $J_{\perp}>0$ the $XY$ model scales to a point on the fixed-point 
line $\Gamma_{\perp}=0$ which approaches the $SU(2)$ end-point at 
$J_{\perp} \rightarrow 0$ and moves along the critical line with increasing
parameter $J_{\perp}$. 

Using (\ref{Gs2order}) and (\ref{Gperp2order}) and applying a similar 
analysis in the case $J_{\parallel} \neq 0$, one easily obtains that a
gapless regime in the spin channel exists for
\begin{equation}\label{RGmu2jparalel}
J_{\perp} \leq \sqrt{2\pi t J_{\parallel}}.   
\end{equation}

\section{The weak-coupling phase diagram}

Let us now consider the weak-coupling ground state phase diagram of the model 
Eq. (\ref{tJmodel}). 


\subsection{Order parameters}

To clarify the symmetry properties of the ground states of the system in 
different sectors of the phase diagram we use the following set of order 
parameters describing the 
short wave--length fluctuations of the {\em site--located} charge density,
\begin{eqnarray}\label{CDWop}
\Delta_{\small CDW} & = &  (-1)^{n} \sum_{\alpha}
c^{\dagger}_{n,\alpha}c_{n,\alpha} \nonumber\\
& \sim &  \sin(\sqrt{2\pi K_{c}}\varphi_{c}) 
\cos( \sqrt{2\pi K_{s}}\varphi_{s}) \, ,
\end{eqnarray}  
the {\em site-located} spin density, 
\begin{eqnarray} 
\Delta_{\small SDW^{z}} & = & (-1)^{n} \sum_{\alpha}\alpha 
c^{\dagger}_{n,\alpha}c_{n,\alpha}\nonumber\\
& \sim &
\cos(\sqrt{2\pi K_{c}}\varphi_{c}) \sin( \sqrt{2\pi K_{s}}\varphi_{s}) 
\label{SDWZop}  \\ 
\Delta_{\small SDW^{x}} & = & (-1)^{n} \sum_{\alpha}c^{\dagger}_{n,\alpha}
c_{n,-\alpha}\nonumber\\
& \sim & 
\cos(\sqrt{2\pi K_{c}}\varphi_{c})\cos(\sqrt{\frac{2\pi}{K_{s}}}\theta_{s}),
\label{SDWXop}\\
\Delta_{\small SDW^{y}} & = & i(-1)^{n} \sum_{\alpha}\alpha
c^{\dagger}_{n,\alpha}c_{n,-\alpha}\nonumber\\
& \sim &
\cos(\sqrt{2\pi K_{c}}\varphi_{c}  )\sin(\sqrt{\frac{2\pi}{K_{s}}}\theta_{s}),
\label{SDWYop}
\end{eqnarray} 
and the short wave--length fluctuations of the {\em bond--located} 
charge density,  
\begin{eqnarray}  
\Delta_{\small dimer} & = &(-1)^{n} \sum_{\alpha}
(c^{\dagger}_{n, \alpha}c_{n+1,\alpha} + H.c.) \nonumber\\
& \sim &
\cos(\sqrt{2\pi K_{c}}\varphi_{c}) \cos( \sqrt{2\pi K_{s}}\varphi_{s}).
\label{DIMERop}
\end{eqnarray} 
In addition we use two superconducting order parameters corresponding to 
singlet ($\Delta_{SS}$) and triplet ($\Delta_{TS}$) superconductivity:
\begin{eqnarray}
\Delta_{SS}(x) &= & R^{\dagger}_{\uparrow}(x)L^{\dagger}_{\downarrow}(x)  
- R^{\dagger}_{\downarrow}(x)L^{\dagger}_{\uparrow}(x) \nonumber\\
& \sim & \exp(i \sqrt{\frac{2\pi}{K_{c}}}\theta_{c}) 
\cos(\sqrt{2 \pi K_{s}}\varphi_{s}),\label{SSop}\\
\Delta_{TS}(x) & = & R^{\dagger}_{\uparrow}(x)L^{\dagger}_{\downarrow}(x)  + 
R^{\dagger}_{\downarrow}(x)L^{\dagger}_{\uparrow}(x) \nonumber\\
& \sim &   \exp(i  \sqrt{\frac{2\pi}{K_{c}}}\theta_{c}) 
\sin(\sqrt{2 \pi K_{s}}\varphi_{s}).\label{TSop}
\end{eqnarray}

\subsection{Phases}

With the results of the previous section for the excitation spectrum 
and the behavior of the corresponding fields 
Eqs.~(\ref{freecorrelations1})--(\ref{orderfields}) we now analyze the ground 
state phase diagram of the model (\ref{tJmodel}) (see Fig. 5 and Fig. 6).

Let us first consider the sector of the phase diagram  corresponding to 
$U+J_{\perp}+\frac{1}{2}J_{\parallel}>0$, characterized by a 
{\bf gap in the charge excitation spectrum}. In this case we obtain the 
following regimes of behavior:

\begin{itemize}
\item {\bf A.} \qquad   $\Delta_{c} \neq 0$,  $\Delta_{s} \neq 0$,  
$\langle\varphi_{c}\rangle = \langle \varphi_{s} \rangle =0$; 
\end{itemize}

This regime corresponds to the appearance of a long-range ordered 
{\em dimerized (Peierls)} phase 
\begin{equation}\label{correlDimer}
\langle \Delta_{\small dimer}(x) \Delta_{\small dimer}(x')\rangle \sim 
{\it constant}
\end{equation}
in the ground state of the model. This phase is realized in the case of {\em 
dominating antiferromagnetic exchange}, in particular for isotropic exchange 
$J_{\parallel}=J_{\perp}=J>2U$. 

\begin{itemize}
\item {\bf B.} $\Delta_{c} \neq 0$,  $\Delta_{s} \neq 0$,  
$\langle\varphi_{c}\rangle = 0$,  
$\langle \varphi_{s} \rangle = \sqrt{\pi/8K_{s}}$;
\end{itemize}

This regime corresponds to the appearance of a long-range ordered 
{\em antiferromagnetic (Neel)} phase
\begin{equation}\label{correlSDWz}
\langle\Delta_{\small SDW^{z}}(x)\Delta_{\small SDW^{z}}(x')\rangle \sim 
{\it constant}
\end{equation}
in the ground state. In this regime the Hubbard model is extended by
incorporating an easy-axis spin exchange interaction.

\begin{figure}
\mbox{\epsfxsize 8.0cm\epsffile{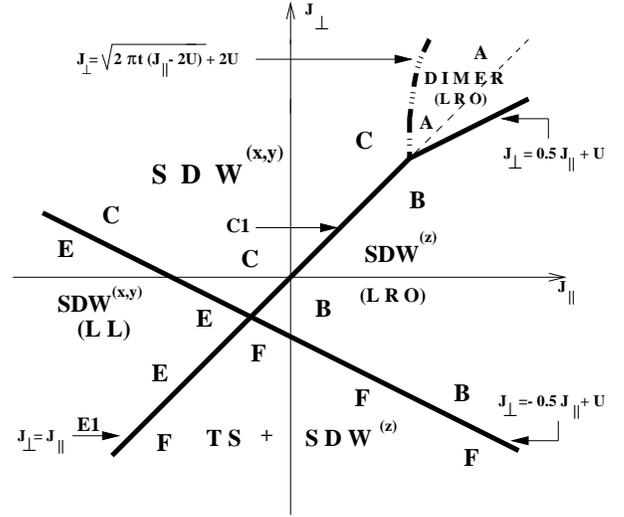}}
\vskip0.5cm
\caption[]{The weak-coupling phase diagram  of the model (\ref{tJmodel}) at 
$U > 0$}
\label{TRSCfig5}
\end{figure} 

\begin{itemize}
\item {\bf C.} \qquad $\Delta_{c} \neq 0$,  
$\langle\varphi_{c}\rangle = 0$, $\Delta_{s} = 0$;
\end{itemize}

The charge excitation spectrum is gapped. Ordering of the field 
$\varphi_{c}$ with vacuum expectation value $\langle \varphi_{c} \rangle =0$ 
leads to a suppression of the $CDW$ and {\em superconducting} correlations. The 
$SDW$ and {\em Peierls} correlations show a power-low decay at large distances. 
The low-energy properties of the gapless spin degrees of freedom are controlled 
by the fixed-point value of the Luttinger liquid parameter $K^{\ast}_{s}$. 

In the $SU(2)$-spin symmetric case
\begin{itemize}
\item {\bf C1.} \qquad $J_{\parallel}=J_{\perp}=J,\qquad  
-\frac{2}{3}U < J < 2U$;
\end{itemize}

$K^{\ast}_{s}=1$ and the {\em Peierls}
and $SDW^{{\it i}}$ $({\it i}= x,y,z)$ correlations show
{\em identical power-law decay} at large distances:
\begin{eqnarray}\label{correlDIMER}
\langle \Delta_{\small dimer}(x) \Delta_{\small dimer}(x')\rangle
&\simeq & \langle\Delta_{\small SDW}(x)\Delta_{\small SDW}(x')\rangle 
 \nonumber\\
& \sim & \left| x - x' \right|^{-1}.
\end{eqnarray}

In the general case of $U(1)$-spin symmetry,
$K^{\ast}_{s}>1$ and the 
{\em ``in-plane''} $SDW^{x,y}$ correlations dominate in the ground state,
\begin{eqnarray}\label{correlSDWxy}
 \langle\Delta_{\small SDW^{x}}(x)\Delta_{\small SDW^{x}}(x')\rangle 
&\simeq & \langle\Delta_{\small SDW^{y}}(x)\Delta_{\small SDW^{y}}(x')\rangle 
 \nonumber\\
& \sim & \left| x - x' \right|^{-1/K^{\ast}_{s}},
\end{eqnarray}
while the {\em Peierls} and $SDW^{z}$ correlations decay faster,
\begin{eqnarray}\label{correldimernLRO}
\langle \Delta_{\small dimer}(x) \Delta_{\small dimer}(x')\rangle
&\simeq & \langle\Delta_{\small SDW}(x)\Delta_{\small SDW}(x')\rangle 
 \nonumber\\
& \sim & \left| x - x' \right|^{-K^{\ast}_{s}}.
\end{eqnarray}
This case corresponds to an extension of the Hubbard model by
incorporating an easy-plane spin exchange interaction.

Let us now consider the sector $U+J_{\perp}+\frac{1}{2}J_{\parallel}<0$ 
in which the charge excitation spectrum is gapless.

The following different regimes are realized in this sector:

\begin{itemize}
\item {\bf D.}\qquad $\Delta_{c} = 0 $, $\Delta_{s} \neq 0$, 
$\langle\varphi_{s}\rangle = 0$;
\end{itemize}

This phase is realized in the case of {\em dominating attractive Hubbard 
interaction}, in particular for isotropic exchange 
$J_{\parallel} = J_{\perp} = J$ at $2U<J<-\frac{2}{3}U$. 

There is a gap in the spin excitation spectrum. The spin field is ordered,
$\langle \varphi_{s} \rangle =0$. Ordering of the field 
$\langle\varphi_{s}\rangle$ leads to a {\em suppression} of the $SDW$ and $TS$ 
fluctuations. The low-energy properties of the 
gapless charge degrees of freedom are controlled by the fixed-point value of 
the Luttinger liquid parameter $K^{\ast}_{c}$. 

In the case of a half-filled band which we are considering here the charge 
degrees of freedom are governed by 
the $SU(2)$-charge symmetry. The fixed-point value of the parameter $K_{c}$ 
(due to the $SU(2)$-charge symmetry) is $K^{\ast}_{c}=1$. The {\em CDW}, 
{\em SS}, and {\em Peierls} correlations show identical power-law decay at 
large distances,
\begin{eqnarray}
\langle \Delta_{\small CDW}(x)\Delta_{\small CDW}(x')\rangle &=& 
\langle \Delta_{\small SS}(x)\Delta_{\small SS}(x')\rangle =\nonumber\\
\langle \Delta_{\small dimer}(x) \Delta_{\small dimer}(x')\rangle &\sim& 
\left| x - x' \right|^{-1}.
\label{correlSU2}
\end{eqnarray}

\begin{itemize}
\item {\bf E.} \qquad   $\Delta_{c} = 0$,  $\Delta_{s} = 0$; 
\end{itemize}

In this case the Luttinger liquid ($LL$) phase is realized.

The charge and the spin channel are gapless. the low-energy behavior of the 
system is controlled by the Luttinger liquid parameters $K^{\ast}_{c}$ and 
$K^{\ast}_{s}$.

In the case of $SU(2)$-invariant spin exchange 
\begin{itemize}
\item {\bf E1.} \qquad $J_{\parallel}=J_{\perp}=J < 
{\em min}\{-\frac{2}{3}U, 2U \}$  
\end{itemize}

$K^{\ast}_{c}=K^{\ast}_{s}=1$ and all correlations show 
$\left| x - x' \right|^{-2}$ decay at large distances. 

In the case of a $U(1)$-symmetric spin channel, the LL phase is realized for
easy-axis ferromagnetic anisotropy. In this case $K^{\ast}_{c}=1$, 
$K^{\ast}_{s}>1$ and in the weak-coupling limit the in-plane antiferromagnetic
correlations dominate
\begin{equation}\label{correlTS+SS2}
\langle \Delta_{\small SDW^{x,y}}(x)\Delta_{\small SDW^{x,y}}(x')\rangle \sim  
\left| x - x' \right|^{-1-1/K_{s}}.
\end{equation}

\begin{itemize}
\item {\bf F.} \qquad $\Delta_{c} = 0$, $\Delta_{s} \neq 0$ 
$\langle\varphi_{s}\rangle = \sqrt{\pi/8K_{s}} $;
\end{itemize}

This sector of the phase diagram is dual to the sector 
{\bf D}. As common in the half-filled band case, the gapless charge 
excitation spectrum opens a possibility for the realization of a
{\em superconducting} instability in the system. Moreover, due to the 
$U(1)$-symmetry of the system, ordering of the $\varphi_{s}$ with vacuum 
expectation value $\langle \varphi_{s} \rangle =\sqrt{\pi/8K_{s}} $ leads to a
suppression of the $CDW$ and $SS$ correlations. In this case the $SDW$ 
and $TS$ fluctuations show identical power-low decay at large distances,
\begin{eqnarray}\label{correlTS-SDW}
\langle \Delta_{\small SDW}(x)\Delta_{\small SDW}(x')\rangle  &=& 
\langle \Delta_{\small TS}(x)\Delta_{\small TS}(x')\rangle \nonumber\\
& \sim & \left| x - x' \right|^{-1},
\end{eqnarray}
and are the {\em dominating instabilities} in the system. This phase is 
realized only for anisotropic spin exchange in the case of strong  
{\em ferromagnetic easy-plane} anisotropy.
\begin{figure}
\mbox{\epsfxsize 8.0cm\epsffile{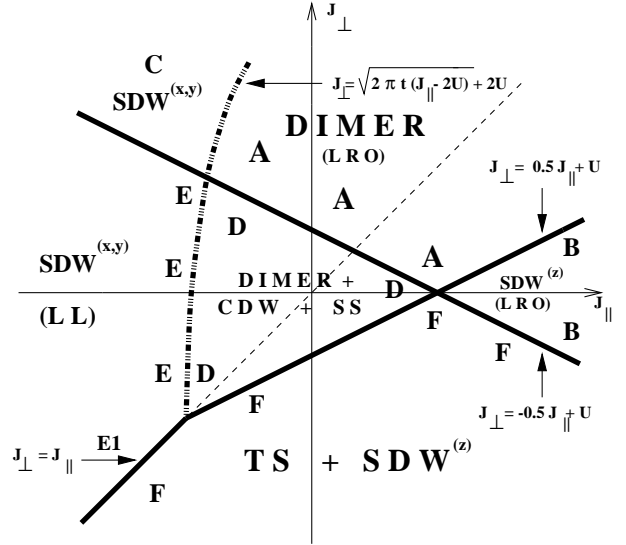}}
\vskip0.5cm
\caption[]{
The weak-coupling phase diagram of the model (\ref{tJmodel})  at $U < 0$}
\label{TRSCfig6}
\end{figure} 

\section{Discussion and summary}

In this paper we have studied the one-dimensional $t-J$ model of 
correlated electrons in the case of a half-filled band. This model describes a 
system of itinerant electrons with spin-exchange interaction between 
electrons on nearest-neighbor sites. We have demonstrated that in 
the case of easy-plane anisotropic ferromagnetic exchange the 
{\em triplet superconducting} and $SDW$ instabilities are the dominating 
instabilities in the system. These instabilities remain dominating 
instabilities in the ground state also for moderate values of the on-site 
Hubbard interaction. We stress that, in 1D this phase can be realized only 
in the case of $U(1)$-spin symmetry. This result is in agreement with recent 
experimental results showing strong easy-plane anisotropy of 
ferromagnetic spin fluctuations in the triplet superconductor $Sr_{2}RuO_{4}$ 
\cite{Trexp5}. We want to stress, that although in this paper we presented
results considering the half-filled band case only, it is obvious that doping 
will split the degeneracy between the $TS$ and $SDW$ phases and will favor the
superconducting instability in the system.

We also demonstrated a strong enhancement of tendencies towards Peierls 
ordering in the electron system caused by an isotropic (or weakly anisotropic) 
antiferromagnetic exchange. We have shown that the half-filled $t-J$ 
model shows a long-range dimerized (Peierls) ordering in the ground state 
in the case of antiferromagnetic exchange.

We also demonstrated the importance of the finite-band effects in this model 
and presented a qualitative description of the transition from the 
band-dominated $TS$+$SDW$ phase into the insulating spin-1/2 magnetic $XY$ 
phase. Detailed numerical studies of the phase diagram for strong exchange 
coupling and for arbitrary filling are in progress and will be published 
elsewhere.
\vskip1.0cm
\centerline{{\bf ACKNOWLEDGMENTS}}
\vskip0.5cm

GJ aknowledges helpful discussions with D. Baeriswyl, C. Dziurzik, A.P. Kampf, 
A.A. Nersesyan and A. Schadschneider. This work was supported by the Deutsche 
Forschungsgemeinschaft in the framework of the research program of the 
Sonderforschungsbereich 341. GJ acnowledges INTAS support under the Grant 
GEORGIA-INTAS No. 97-1340.


\begin{thebibliography}{99}
\bibitem{Note}{Here we use the analogy between the itinerant-$XY$ and the 
pair-hopping model. The one spin component electron-hole transformation 
$c_{n,\uparrow} \rightarrow c_{n,\uparrow}, 
c_{n,\downarrow}  \rightarrow  (-1)^{n}c^{\dagger}_{n,\downarrow}$ 
interchanges the charge and spin degrees of freedom, and 
maps the transverse exchange part into the pair-hopping coupling 
$\frac{1}{2}J_{\perp}(S^{+}_{n}S^{-}_{n+1} + h.c.)\rightarrow 
-\frac{1}{2}J_{\perp}(c^{\dagger}_{n,\uparrow}c^{\dagger}_{n,\downarrow}
c_{n+1,\downarrow} c_{n+1,\uparrow} + h.c.)$. In the case of a half-filled 
band a transition from the insulating into an $\eta$-superconducting phase 
takes place at $-\frac{1}{2}J_{\perp} \sim 2t$ in the pair-hopping model. 
This transition is first order (level crossing) and at the 
transition the charge gap closes and a spin gap opens \cite{SA}}.
\end{thebibliography}
\end{document}